\documentclass{osa-article}
\journal{osac}

\articletype{Research Article}

\begin{document}

\title{A vertically-loaded diamond microdisk resonator spin-photon interface}




\author{Yuqin Duan\authormark{1,2}, Kevin C. Chen\authormark{1,2}, Dirk R. Englund\authormark{1,2,$\dagger$}, and Matthew E. Trusheim\authormark{1,3,*}}

\address{\authormark{1}Research Laboratory of Electronics, M.I.T., Cambridge, MA, 02139, USA\\
\authormark{2} Department of Electrical Engineering and Computer Science, M.I.T., Cambridge, MA 02139, USA\\
\authormark{3}U.S. Army Research Laboratory, Sensors and Electron Devices Directorate, Adelphi, MD 20783, USA}

\email{\authormark{$\dagger$}englund@mit.edu}
\email{\authormark{*}mtrush@mit.edu}



\begin{abstract}
We propose and optimize a vertically-loaded diamond microdisk resonator (VLDMoRt) coupled to a nitrogen-vacancy (NV) center in diamond for efficient collection of zero-phonon-line emission into low numerical aperture (NA) free-space modes. The VLDMoRt achieves a Purcell enhancement of 172 with $39\%$ of the emitted light collected within a NA of 0.6, leading to a total external spin-photon collection efficiency of 0.33. As the design is compatible with established nanofabrication techniques and couples to low-NA modes accessible by cryogenic free-space optical systems, it is a promising platform for efficient spin-photon interfaces based on diamond quantum emitters. 
\end{abstract}

\section{Introduction}
A central challenge in quantum information processing is the efficient entanglement of long-lived quantum memories. Quantum emitters in diamond such as the nitrogen-vacancy (NV) center\cite{Doherty2013-ay,Abobeih2018-ae,De_Lange2010-un} and the group-IV emitters \cite{Bradac2019-qa,Rogers2014-ea,Zhang2018-uq} are promising candidates for scalable entanglement schemes\cite{Munro2010-ua}, as they exhibit excellent optical \cite{Trusheim2020, Rogers2014-ea} and spin coherence properties\cite{Abobeih2018-ae,Childress281,Petr2017} . However, their optical transitions spontaneously emit into many spatial and spectral modes, such as phonon sideband (PSB) transitions, thereby limiting the spin-photon entanglement generation rate\cite{Humphreys2018-ik}. Photonic nanostructures have been explored to overcome this issue by engineering the optical modes resonant with the quantum emitter. Increased optical density of states (Purcell enhancement) in optical cavities can enable efficient single-mode coupling as spontaneous emission is directed into the cavity mode. Additionally, the cavity mode can be engineered for outcoupling to waveguides or specific free-space modes for measurement or transmission to a larger network\cite{Burek2017-fu,Robledo2011-ih}. Several photonic devices have made use of these effects to improve the performance of quantum emitter spin-photon interfaces, including one- and two-dimensional photonic crystal cavities\cite{Burek2014-ek,Wan2018-ok,Mouradian2017-oi}, ring resonators\cite{Hausmann2014-zm}, fiber-based Fabry-Perot cavities\cite{Ruf2021-yf,Janitz2015-bz}, and circular grating outcouplers\cite{Li2015-ru}. While previous work has independently achieved high emitter-cavity cooperativities (e.g. $C \gg 1$ in nanocavities) or efficient coupling to free-space ($90\%$  collection in optical antennas), simultaneous optimization of both parameters for maximal efficiency has not been fully explored. Here we introduce a vertically-loaded diamond microdisk resonator (VLDMoRt) that simultaneously achieves efficient NV emission into the cavity mode via Purcell enhancement, and efficient collection into a low numerical aperture (NA) free-space mode via an optimized perturbative grating. The VLDMoRt reaches a Purcell enhancement of 172 at the NV zero phonon transition (637~nm), with $39\%$ of emitted field intensity within a far-field NA = 0.6. This results in a total spin-photon entanglement efficiency of $0.33$ using a geometry compatible with monolithic diamond fabrication based on quasi-isotropic etching\cite{Wan2018-ok,Mouradian2017-oi,Mitchell2019-iv}.

Current quantum networking schemes are limited by the entanglement efficiency $\eta$ that parameterizes the probability per attempt of collecting a spin-entangled photon into the larger optical apparatus (i.e. the quantum network link)\cite{Humphreys2018-ik}. Here we consider spin-photon entanglement generated via spin-selective photon emission from a quantum emitter, in particular the NV center in diamond. This process is inefficient ($\eta \ll 1$) due to preferential emission into the PSB\cite{Doherty2013-ay} and small overlap between the spatial emission profile of the NV dipole and the collection optics. To quantify these separate mechanisms, we break $\eta$ into two components $\eta = \eta_1 \eta_2$, where $\eta_1$ quantifies the efficiency of emission into the spin-dependent, cavity-coupled zero-phonon line (ZPL) and $\eta_2$ is the far-field collection efficiency based on spatial overlap of the emitted mode with the collection aperture. The spectral efficiency $\eta_1$ is defined as\cite{Li2015-sm}:

\begin{equation}
    \eta_1 = \frac{F}{F+(\frac{\Gamma_{\text{total},0}}{\Gamma_{\text{ZPL},0}}-1)}
\end{equation}
where $F = 3/4\pi^2(Q/V)$ is the Purcell factor of the ZPL within a cavity of quality factor $Q$ and normalized mode volume $V$, $\Gamma_{\text{total},0}$ is the emitter's total emission rate in the absence of cavity enhancement, and $\Gamma_{\text{ZPL},0}$ is the unmodified emission rate of the ZPL. Under off-resonant pumping, the NV center cycles between the ground and the excited states, leading to fluorescence emission across a broad spectrum consisting of a PSB (97\% of emission) and a ZPL peak at 637~nm (3\% of emission), giving $\Gamma_{\text{total},0}/\Gamma_{\text{ZPL},0}-1 \approx 30$ (Fig.~1a). As a result, the Purcell factor is linearly related to $\eta_1$ in the regime of $F \ll 30$, at which point the gains in efficiency begin to saturate.

The collection efficiency into a lens of a given numerical aperture (NA), or equivalently within a far-field divergence angle of $\theta_c =\sin^{-1}(\text{NA}) $, is
\begin{equation}
    \eta_2 = \frac{\int_{0}^{2\pi}\int_{0}^{\theta_c} |\vec{E}|^2 \,\sin\theta\, d\theta \,d\psi \ }{\int_{0}^{2\pi}\int_0^{\pi} \lvert \vec{E} \rvert^2 \sin\theta\,d\theta \,d\psi \ }
\end{equation}
where $\lvert \vec{E} \rvert^2=\left|\vec{E}(\vec{r})e^{i \omega t}\right|^2=\left|\vec{E}(\vec{r})\right|^2$ is evaluated in the far-field at $|\vec{r}|\gg \lambda$ for a monochromatic field at angular frequency $\omega$ and wavelength $\lambda=2\pi c/ \omega$. 





\begin{figure}[h!]
\centering\includegraphics[width=12cm]{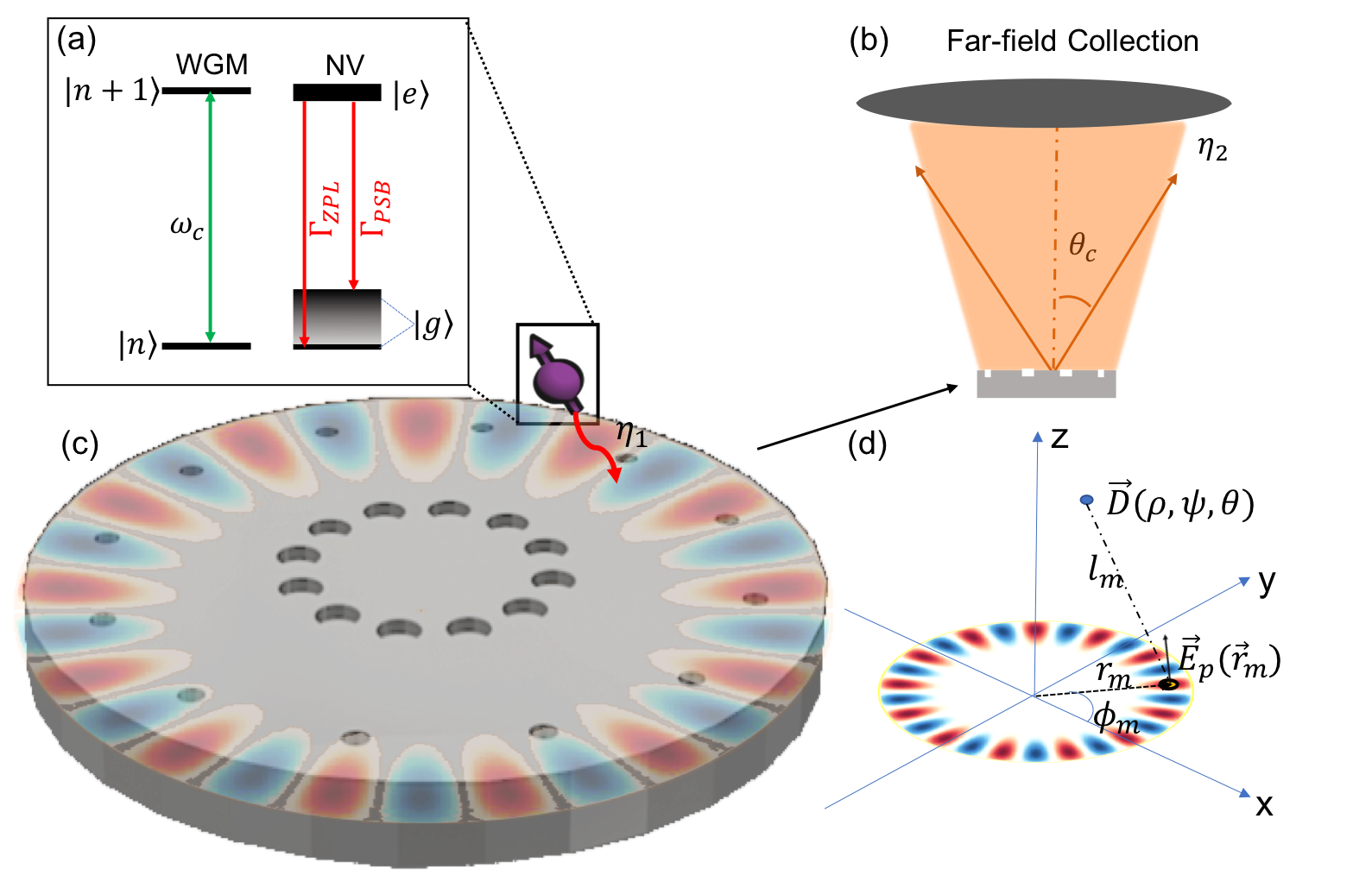} 
\caption{VLDMoRt concept. An NV center is coupled to a WGM resonator with efficiency $\eta_1$, which in turn emits into the far-field collection mode with efficiency $\eta_2$ through an embedded perturbative grating. (a) The WGM is on-resonance with the NV center's ZPL. (b) Side view of VLDMoRt coupling to free space with emission angle $\theta_c$. (c) The VLDMoRt design with a WGM excited. Red and blue represents the WGM near-field, and concentric, periodic pertubative gratings at two radii are shown. (d) Schematic diagram of the scattering model for far-field intensity analysis.}
\end{figure}

Maximizing $\eta$ requires optimizing the tradeoff between $\eta_1$, which is maximized in the case of zero radiative losses, and $\eta_2$, which is maximized when radiative coupling to the desired free-space mode is the dominant loss mechanism. As $\eta_1$ saturates with Purcell enhancement, an optimal design can be reached with low $Q/V \sim 10^2$ in comparison to Purcell-optimized nanocavities or bowtie antennas that achieve $Q/V$ exceeding $10^4$\cite{Evans662,Karamlou:18}. Whispering gallery mode (WGM) resonators offer high quality factors at the cost of high mode volume, which satisfies the $Q/V$ requirements of this application while offering the opportunity to shape the far-field emission by engineering the radiation of the near-field over multiple spatial periods. In particular, our proposed VLDMoRt design achieves efficient free-space collection by use of a numerically-optimized, partially-etched grating structure embedded within a WGM, as described below.

\section{VlDMoRt Design}

The VLDMoRt geometry is summarized in Fig.~1. It consists of a WGM resonator coupled to an NV center along with a periodic grating outcoupler (Fig.~1c). A microdisk of radius $R$ has mode solutions indexed by azimuthal number $P$, corresponding to resonances at wavelengths $\lambda = \frac{2\pi n_{\text{eff}} R}{P}$ and field profiles $\vec{E}_{P} \propto e^{jP\phi}$, where $n_{\text{eff}}$ is the effective refractive index \cite{Wang1997}. In the limit of small microdisk thickness $h < \frac{\lambda}{2n_{\text{eff}}}$ there exists only a single confined transverse mode for each index $P$. The disk supports modes of both TE ($E_z = 0$) and TM ($E_{r}, E_{\phi}  = 0$) polarization, as well as degenerate modes propagating with $ e^{-jP\phi}$ and $e^{jP\phi}$. We consider standing wave solutions that are the superposition of $\pm P$, giving $\vec{E}_P \propto \cos(2 \pi P\phi)$. To obtain high spectral efficiency $\eta_1$ by Purcell enhancement, we choose a disk radius $R$ and thickness $h$ such that a single cavity mode $E_P$ is resonant with the ZPL transition.

To increase the spatial collection efficiency $\eta_2$, we design a periodic grating to couple the WGM into a low-NA free space mode. We approximate the scattering from a sub-wavelength refractive index perturbation located at a point $\vec{r}_m$ as dipole emission with a corresponding dipole moment $\vec{p}$. The scattered field at location $\vec{D}$ in this approximation is\cite{griffiths2017introduction}


\begin{equation}
\label{eqn:radiation field}
    \vec{E}_m = \frac{1}{4\pi\epsilon_0}\Big\{\frac{\omega^2}{c^2l_m}(\widehat{l}_m\times\vec{p})\times\widehat{l}_m + (\frac{1}{{l_m}^3}-\frac{i\omega}{c{l_m}^2})(3\widehat{l}_m[\widehat{l}_m\cdot\vec{p}]-\vec{p})\Big\}e^{\frac{j\omega {l_m}}{c}}e^{-i\omega t}
\end{equation}
where $l_m\widehat{l}_m = \vec{D}-\vec{r}_m$ and $\omega$ is the angular frequency of the dipole (see Fig.~1d). We further take the scattering to originate from the local WGM electric field $\vec{E}_P$, such that the effective dipole at the point of scattering is $\vec{P} = A\vec{E}_P(\vec{r}_m)$ with $A$ being a polarizability dependent on the geometry of the scatterer. In the far-field $\frac{\omega l_m}{c} \gg 1$, the second and third terms proportional to $\omega/c{l_m}^2, 1/{l_m}^3$ are small compared to the term $\propto \frac{\omega^2}{c^2l_m}$, and we neglect them in the remainder of the analysis. The far-field electric profile $\vec{E}(\vec{D})$ originating from a total of $G$ subwavelength scatterers located at points $\vec{r}_m$ is then\cite{von1995dielectrics}
\begin{equation} \label{eqn:total_far_field}
    \vec{E}(\vec{D}) = \frac{A\omega^2}{4\pi\epsilon_0 c^2}\sum_{m=1}^{G}e^{jkl_m} \frac{1}{l_m} (\widehat{l}_m\times\vec{E}_{P}(\vec{r}_m))\times\widehat{l}_m
\end{equation}
where $k = 2\pi/\lambda$ is the free-space wavevector, and the polarizability $A$ is taken to be uniform for all scatterers $m$. We consider the case of evenly-spaced scatterers forming a grating along the resonator circumference at a fixed radius $r_s$ and corresponding angles $\phi_m = 2\pi m/G$\cite{Zhu2013-jf}. As both the WGM near-field and the pertubations are periodic, the amplitude of the electric field at the pertubations is periodic with the difference frequency $L = P - G$ between the two. The field at the location of the scatterers is then $\vec{E}_{P}(\vec{r}_m)= E_{r_s}\cos(L\phi_m) \widehat{E}_P(\phi_m)$, where $E_{r_s}$ is the maximum WGM amplitude at the scattering radius. We have left the mode polarization $\widehat{E}_P(\phi_m)$ as arbitrary to be discussed below.
 
 
For collection at low angles in the far field $\widehat{l}_m\cong \widehat{z}_m$, Eq.~\ref{eqn:total_far_field} simplifies to:
\begin{equation}
    \vec{E}(\vec{D}) =  \frac{A\omega^2}{4\pi\epsilon_0 c^2} E_{r_s}\sum_{m=1}^{G}\frac{e^{jkl_m}}{l_m} \cos(L\phi_m)\widehat{E}_p({\phi}_m)
\end{equation}

 Here the difference in spatial periodicity $L$ determines the far-field interference pattern. We let $l_m$ be a constant for all $m$, and when $L = 0$  ($P = G$), the scattered field of each individual scatterer is in-phase (Fig.~2d) and $\cos(L\phi_m) = 1$. However, the sum over the polarization $\sum\widehat{E}_p(\phi_m)$ is zero for the cases of $\widehat{E}_p(\phi_m) = \widehat{r}_m$ (radial polarization) and $\widehat{E}_p(\phi_m) = \widehat{\phi}_m$ (azimuthal polarization). This results in destructive interference at low emission angles for perturbations matching with the WGM periodicity. With $L = 1$ (Fig.~2g), the cosine term inverts sign ($\pi$ phase shift) for pertubations on opposite sides of the microdisk,  $\cos(\phi_m) = -\cos(\phi_{G/2+m})$ as $\phi_{G/2+m} = \phi_m + \pi$, and field vector likewise inverts, $ \widehat{E}_p(\vec{r}_m) = -\widehat{E}_p(\vec{r}_{G/2+m})$, causing constructive interference at low angles. The interference profile of TE-like modes can therefore be shaped depending on the grating periodicity, while TM-like (out-of-plane $\widehat{z}$ polarized) modes cannot. Intuitively, the offset between the perturbation and mode produces scattering that is out-of-phase on opposite sides of the disk, leading to a dipole-like radiation pattern\cite{Mirzapourbeinekalaye2020-br}. 
 
 We note that this grating configuration results in pertubations spaced by $2r_s$ across the center of the disk, effectively creating a double-slit-like interference pattern\cite{Saleh1991}. For $r_s > \lambda/2$, higher-order ($\theta_c > 0$) interference maxima will exist in addition to the low-angle maximum described above. Small resonators ($r_s < \lambda/2$), or the addition of a second row of perturbations, can eliminate these higher-order interference maxima, potentially channeling all emission into low NA.
 
 
\section{Simulation}

\begin{figure}[h!]
\centering\includegraphics[width=12cm]{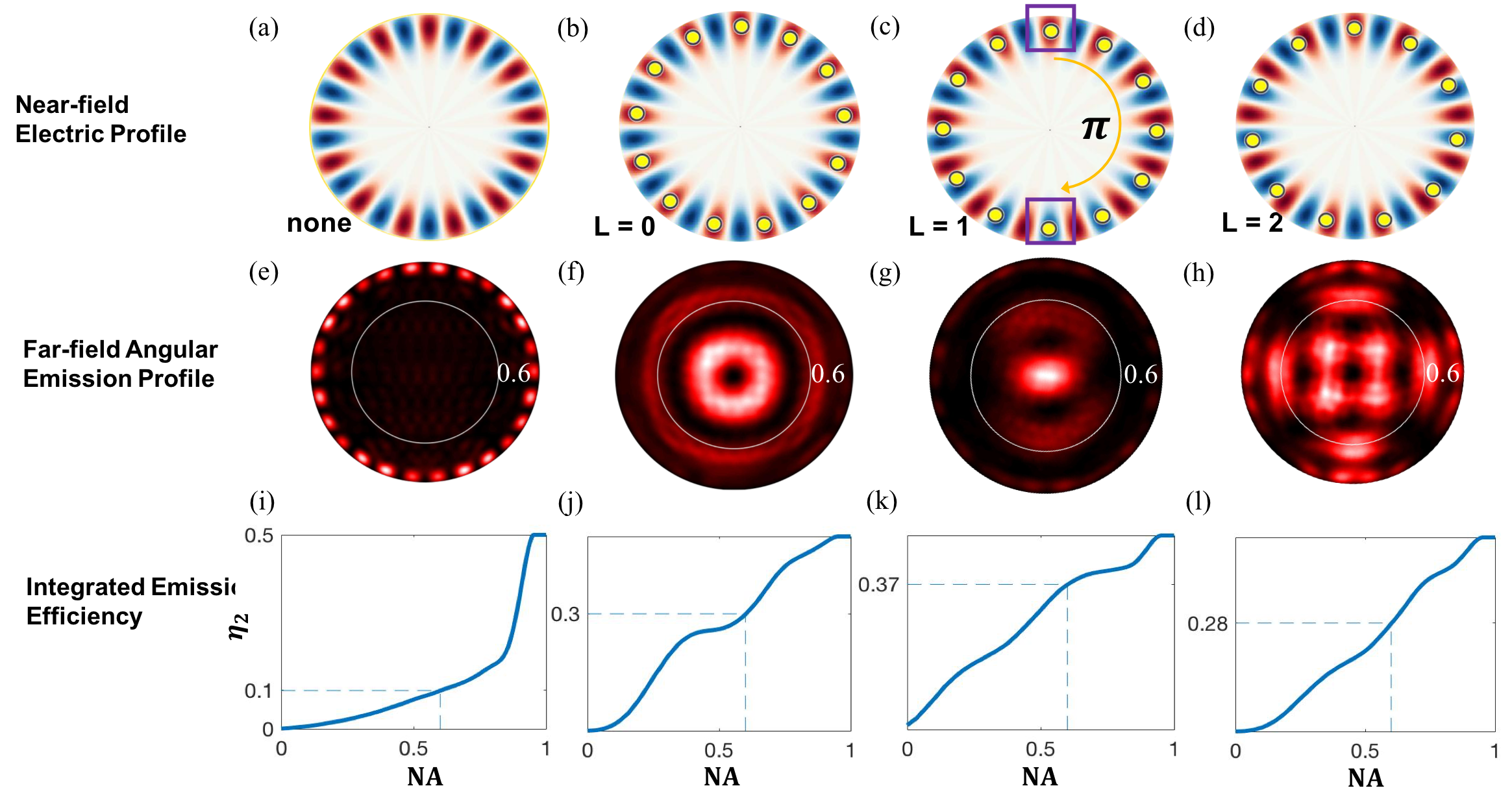}
\caption{Far-field emission profiles. The first row shows the near-field electric profile of a $P = 13$ with perturbations indicated by yellow circles for the four considered geometries (no grating and $L = 0, 1, 2$). The second row shows the far-field angular emission profile with the white circle indicating NA = 0.6. The third row shows the integrated far-field emission intensity within certain NA, and $\eta_{2}$ at NA = 0.6 is marked on each plot. Each column refers to a dielectric geometry as labelled. The boxes in (c) highlight a $\pi$ shift between scatterers of opposite positions in a microdisk.}
\end{figure}

We analyze the performance of the VLDMoRt using finite-difference time-domain (FDTD) simulations of the electromagnetic field (Lumerical Inc.). The diamond WGM resonator has a material index $n = 2.41$ and is surrounded by air ($n=1$). A dipole point source with an emission spectrum centered at 637~nm and orientation in the (111) plane\cite{Epstein2005-te} simulates an NV center. First, we evaluate the predicted far-field emission profiles of different grating orders $L$ based on a test model of a $P = 13$ WGM. The gratings consist of $G$ fully-etched holes with radii $r = 30$~nm located periodically around the disk at a radial position $r_s$ matching the maximum of the WGM near-field $E_P$ (see Fig.~2(b-d)). 

Fig.~2(e-h) shows the calculated far-field emission angular distribution and Fig.~2(i-l) the integrated angular intensity for a bare WGM and for gratings with $L = 0,1,2$. We quantify the integrated intensity as $\eta_2$, which is the fraction of emission collected in a cone defined by a given NA (Eq.~(2)). Without the perturbative gratings, the far-field emission concentrates at NA $> 0.87$. As predicted from Eq.~(4), the $L = 0$ grating results in an anti-node at low angles, forming a ``donut'' beam. Compared to the WGM without a grating, however, the $L = 0$ perturbations improve emission with $\eta_{2} = 0.3$ at NA = 0.6. The $L = 1$ design maximizes emission into low NA as shown in Fig.~2h. As predicted by theoretical analysis, this design results in a large fraction of emission into low angles with $\eta_{2} = 0.37$ and a near-Gaussian mode profile near $\theta=0$. Finally, a higher-order $L = 2$ results in a donut-like mode with additional quadrupolar lobes. The efficiency at low angles is reduced in this case with $\eta_{2} = 0.27$ at NA = 0.6. 

We note that all systems emit only $50\%$ into the upper half-plane ($z > 0$) due to the unbroken vertical symmetry. We also note that the far-field profile in Fig.~2f-h has secondary lobes at higher NA, which lowers the overlap with the preferred low-NA Gaussian profile. As discussed above, this is expected for a grating with $r_s > \lambda/2$, which is the case for this disk ($r_s = 0.729$~$\upmu$m). To further improve $\eta_2$, we consider partially etched gratings which break the vertical symmetry and enhance emission into the upper half-plane. We additionally consider adding a second row of perturbations at the given angles $\phi_m$ to shape higher-order constructive interference in the radial direction.

\begin{figure}[h!]
\centering\includegraphics[width=12cm]{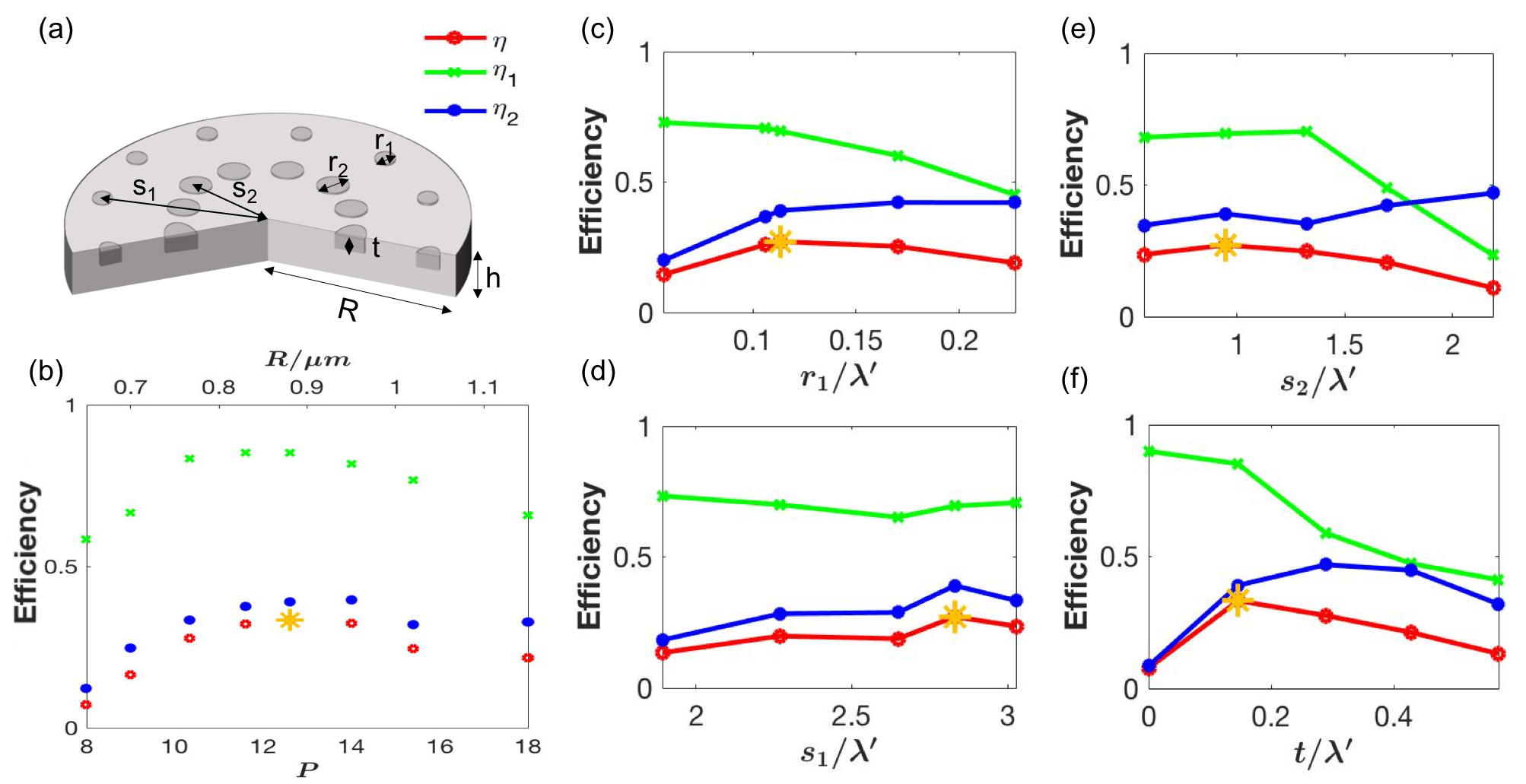}
\caption{Parameter optimization. (a) The dielectric profile defined by the microdisk of radius $R$, thickness $h$, grating radii $r_1$ and $r_2$ at positions $s_1$ and $s_2$, and etched depth $t$. (b) Optimized grating parameters in the set of R corresponding to different resonance P. The yellow star indicates the maximum $\eta = 0.33$ when $P = 13$. (c-f) The dependence of $\eta_1$, $\eta_2$, and $\eta$ on $r_1$, $t$, $s_1$, $s_2$ normalized by $\lambda' = \lambda/n$ separately. The red curve is the overall efficiency $\eta$, while blue and green curve plot $\eta_1$ and $\eta_2$ respectively. The yellow star indicates the optimized solution.}
\end{figure}

Beginning with the $L=1$ design described by the analysis above, we numerically optimize the VLDMoRt for the figure of merit $\eta$. Fig.~3a summarizes the initial geometry.
We consider the following parameters as the optimization variables (see Table 1): the WGM disk radius $R$, disk thickness $h$, outer grating pertubation radius $r_1$ and  radial location from the center $s_1$, and inner grating pertubation radius $r_2$ and radial location $s_2$. We initially set $h = \lambda/2n$ to confine a single vertical mode. Because changes in $R$ result in changing the mode order $P$ resonant with the NV ZPL, we first identify the set of radii $R_P$ corresponding to different resonant modes $P$, and subsequently optimize the grating parameters for each fixed $R_P$ using the built-in particle swarm method in 3D FDTD (Lumerical). The results are shown in Fig.~3b. The maximum $\eta$ is reached using a $P = 13$ WGM, with the optimized parameters shown in Table 1. 

 


\begin{table}[h!]
\begin{center}
\caption{Optimized VLDMoRt Structure}
\begin{tabular}{|l|l|l|l|l|l|l|l|}
\hline
Optimized Result     & $R$ & $h$    & $r_1$ & $r_2$ & $s_1$ & $s_2$ & $t$    \\ \hline
Units of $\lambda'$ & 3.33                     & 0.57 & 0.11 & 0.19 & 2.83 & 0.95 & 0.15 \\ \hline
\end{tabular}
\end{center}
\end{table}

To understand the relationship between the parameters and $\eta$, we analyze the change in efficiency with deviations from the optimized parameters in Fig.~3b-f. The grating design offers trade-offs between $\eta_1$ and $\eta_2$. Both $r_1$ and $s_1$ influence the scattering amplitude from the WGM, trading off coupling to the desired far-field mode with resonator loss and therefore Purcell factor. This is shown clearly in Figure 3c, where increasing $r_1$ above the optimized $r_1 = 0.11$ leads to an increase in $\eta_2$ at the cost of decreasing $\eta_1$, resulting in an overall loss of efficiency $\eta$. Similar to $r_1$, $s_1$ also influences the scattering amplitude, with a maximum at the position of highest near-field intensity $E_P$. 

The design of the second ring of gratings has a less obvious trade-off between quality factor and emission directivity. Fig.~3d shows the relationship between $s_2$ and $\eta$, with $s_2 = 0.95\lambda'$ being the optimized solution. Since the near-field intensity at the inner ring is small compared with the outer ring, maintaining comparable scattering amplitudes between the two rows requires $r_2>r_1$. Indeed, we find that $r_2$ is maximized at $0.19\lambda'$, where the inner ring's scattering field constructively interferes with that of the outer ring.

Increasing the etch depth $t$ also increases scattering. Unlike $r_1$ and $s_1$, however, changing $t$ breaks the vertical symmetry of the VLDMoRt, thereby producing a directional emission into (or away from) the upper half-space. These effects are maximized for $\eta$ at $t = 0.15\lambda'$, where $56\%$ of emission is directed upwards for a $\eta_2 = 0.39$. 

\section{Discussion}
The optimized VLDMoRt achieves high $\eta_1 = 0.77$, but relatively lower $\eta_2 = 0.39$ primarily limited by emission into the lower half plane (>40$\%$ towards the substrate). One potential solution is to add a reflective layer under the device to improve directional radiation\cite{Chen2017-xg,Li21}. In practice, there will be a diamond substrate under the WGM, which we did not consider in the simulation. By appropriately choosing the separation distance between the underside of the WGM resonator and the back reflector, leakage into the substrate can be reduced via destructive interference. Diamond, as a dielectric material, is not a perfect reflector, so only a fraction of leakage to the lower plane will be directed upwards. Depositing an additional metallic layer on the substrate will increase the reflectivity, resulting in an maximal improvement of $\eta_2$ by a factor of 1.8 corresponding to $\eta = 0.54$. 

In this work, we have considered the collection efficiency of an NV center's emission into a low-NA cryogenic objective as the figure of merit. However, a different figure of merit may be desirable depending on the external optics. One example is to directly couple the resonator mode into an optical fiber. In this case, the goal would be matching the far-field emission profile to a Gaussian fiber mode. We note that the $L=1$ geometry already achieves good overlap with a Gaussian mode, and expect that this could be improved with direct optimization of the grating structure for this particular figure of merit. Other optimization procedures can be used, such as inverse design~\cite{Dory2019-gj,chen2020design}, to numerically evaluate performance based on the analytical direction described here. 

VLDMoRt-type grating structures can also be designed to enhance $\eta$ of diamond group-IV emitters or other solid-state defects such as SiC defects\cite{Christle2015-pq}, rare-earth ions\cite{Kolesov2012-ua} and quantum dots\cite{Morse2017-ns,Brooks2021-kz}. The silicon-vacancy (SiV) center in diamond is promising because its spontaneous emission is highly concentrated in the ZPL\cite{Muller2014-ch}. In this case, $\eta$ can be enhanced to $0.41$ which is a $30\%$ improvement over that of the NV center with the same design. Material platforms with demonstrated high-quality WGM resonators are also of particular interest\cite{Cai2000-tv}. Silicon, for example, has a high refractive index that enables high mode confinement but prevents efficient outcoupling, so applications using vertical far-field collection could benefit from the VLDMoRT design.



\section{Conclusion}
This work proposed a new design, VLDMoRt, which radiates directionally into free-space for efficient collection of emission from diamond quantum emitters. Analysis based on scattering theory identified a candidate design based on a $L = 1$ perturbative grating, and numerical optimization in FDTD simulations resulted in an spin-photon entanglement efficiency $\eta = 0.33$. The VLDMoRt can be fabricated with existing quasi-isotropic techniques~\cite{Wan2018-ok}, potentially achieving scalable arrays of efficiently free-space coupled and cavity-enhanced quantum emitters.

\begin{backmatter}
\bmsection{Acknowledgments}
The authors thank R. Camacho for insightful comments. This material is based upon work supported by the U.S. Department of Energy (DOE), Office of Science, Office of Chicago under Award Number DE-SC0020115, the National Science Foundation (NSF) Center for Integrated Quantum Materials (CIQM). Y. D. acknowledges support from the Edwin Webster Fellowship. K. C. C. acknowledges funding support by the National Science Foundation Graduate Research Fellowships Program (GRFP) and the MITRE Corporation Moonshot
program.  M. E. T. acknowledges support through the Army Research Laboratory ENIAC Distinguished Postdoctoral Fellowship.

\bmsection{Disclosures}
The authors declare no conflicts of interest.
\end{backmatter}

\bibliography{VLDM}

\end{document}